# Optical investigations on electronic structure changes related to the metal-insulator transition in VO$_2$ film

C. H. Koo, J. S. Lee, M. W. Kim, Y. J. Chang, and T. W. Noh*
*Research Center for Oxide Electronics and School of Physics* and
*Seoul National University, Seoul 151-747, Korea*

J. H. Jung
*Department of Physics, Inha University, Incheon 402-751, Korea*

B. G. Chae and H.-T. Kim
*Telecom. Basic Research Laboratory, ETRI, Daejon 305-350, Korea*

We investigated optical absorption coefficient spectra of an epitaxial VO$_2$ film in wide photon energy (0.5 - 5.0 eV) and temperature (100 - 380 K) regions. In its insulating phase, we observed two *d-d* transition peaks around 1.3 eV and 2.7 eV and a charge transfer peak around 4.0 eV. As temperature goes above the metal-insulator transition temperature near 340 K, a large portion of the spectral weight of the peak around 4.0 eV becomes redistributed and a Drude-like peak appears. We initially applied the band picture to explain the details of the spectral weight changes, especially the temperature-dependent shift at 2.7 eV, but failed. To check whether the spectral changes are optical signatures of the electron-electron correlation effects, we applied the Hubbard model which takes into account orbital degeneracy. This orbitally degenerate Hubbard model could explain the details of the temperature-dependent peak shifts quite well. In addition, from the peak assignments based on the orbitally degenerate Hubbard model, we could obtain the values of $U + \Delta$ (~ 3.4 eV) and $J_H$ (~ 0.7 eV), where $U$, $\Delta$, and $J_H$ are the on-site Coulomb repulsion energy, the crystal field splitting between the $t_{2g}$ bands, and the Hund's rule exchange energy, respectively. Our spectroscopic studies indicate that the electron-electron correlation could play an important role in the metal-insulator transition of VO$_2$.

73.61.-r, 78.20.-e, 78.40.-q, 78.66.-w

*e-mail address: twnoh@snu.ac.kr

# I. INTRODUCTION

Vanadium dioxide (VO$_2$) has long been widely investigated due to its unique physical properties and possible applications. It shows an abrupt first-order metal-insulator transition (MIT), concomitant with a tetragonal-monoclinic structure change near 340 K.[1] Due to its large change of electric conductivity near room temperature, several researchers have tried to use this material for novel applications, such as field effect transistor and ultra-violet detector.[2,3]

Although this material had been extensively studied since the 1950s,[1] there is still no consensus on the dominant mechanism of its MIT. Goodenough pointed out the importance of lattice instability during the phase transition, *i.e.* the dimerization of the vanadium ions due to the tetragonal-monoclinic structural change.[4] This scenario was supported by Raman scattering[5] and x-ray diffraction experiments,[6] which focused on the structural change. Also, the band structure[7] and phonon dispersion calculations[8] suggested the possible existence of charge-density wave or lattice instability in the metallic phase. On the other hand, Zylbersztejn and Mott pointed out the importance of electron correlations during the phase transition, especially for electrons with the $t_{2g}$ symmetry.[9] This scenario was supported by electron paramagnetic resonance (EPR) and nuclear magnetic resonance (NMR) experiments.[10,11] Even with recent experimental tools and theoretical calculations, a clear explanation for the MIT mechanism is still lacking. For example, Cavalleri *et al.* investigated the femtosecond structural dynamics of VO$_2$ using time-domain ultrafast spectroscopy and initially proposed the importance of the electron-electron correlation.[12] Later, with increase of experimental resolution, they suggested the electron-phonon interaction could be as important as the electron-electron correlation.[13] It is clear that it would be quite difficult to elucidate the underlying MIT mechanism of VO$_2$.

Optical spectroscopy has proved to be a powerful tool for investigating MIT in numerous strongly correlated electron systems. Numerous researchers had tried to address electronic structure changes due to the MIT of VO$_2$, especially near the Fermi level, using optical spectroscopy.[14,15,16] Up to this point, most optical studies on VO$_2$ have been performed based on the band picture, where the electron-electron correlation effects cannot be taken into account properly. However, as it will be shown in Section IV, the optical spectra of VO$_2$ have a multiple peak structure and/or broad spectral feature, which cannot be easily explained by the conventional band picture. Therefore, it would be highly desirable to perform optical spectroscopic studies on VO$_2$ again and to check whether its spectra can be understood in terms of the electron correlation effects.

To explain optical properties of transition metal oxides, the single-band Hubbard model has been commonly used.[17] However, in this model, there will be only one optical transition from the lower Hubbard band to the upper one, so it also fails to explain the multiple *d-d* transition peaks observed in optical spectra of numerous transition metal oxides.[18,19] To overcome this shortcoming, we introduced the orbitally degenerate Hubbard model (ODHM), which explicitly takes into account the orbital degeneracy, and applied the model to explain the optical spectra of some metal oxides, such as molybdates[20] and ruthenates.[21] This new approach can also explain multi-peak structures of *d-d* transitions and their peak positions quantitatively in the early transition metal oxides, such as La*M*O$_3$ (*M* = Ti, V, and Cr).[22] It should be noted that earlier optical studies tended to provide much smaller values of the on-site Coulomb repulsion energy $U$ for transition metal oxides. However, the ODHM can provide values of physical parameters, such as $U$ and the Hund's rule exchange energy $J_H$, which are consistent with those from other experimental measurements.[22] Recently, this model could elucidate that the controversial 2.0 eV peak of LaMnO$_3$ should come from the interatomic transition of the $e_g$ electrons with the *A*-type spin/orbital ordered correlation.[23,24] More recently, we also found that the 2.0 eV peak of *R*MnO$_3$ (*R* = La, Pr, Nd, Gd, and Tb) becomes strongly suppressed with the substitution of the small size rare earth ions, and explained this intriguing behavior based on the ODHM.[25] The recent success of the ODHM suggests that this model could be a new plausible starting point to check the existence of correlation signatures in the VO$_2$ optical spectra and their meanings in the MIT mechanisms.

In this paper, we investigate the optical absorption coefficient spectra, α(ω), of a VO$_2$ film in a wide energy range (0.5 - 5.0 eV) with a systematic variation of temperature (100 - 380 K). We observed a strong *p-d* transition peak around 4.0 eV and two *d-d* transitions near 1.3 eV and 2.7 eV. We found that the multiple *d-d* transitions cannot be explained in terms of the widely accepted band picture. We extended the ODHM to a case where the energy levels of $t_{2g}$ electrons are split by the crystal field splitting $\Delta$, and successfully applied it to explain α(ω) of the VO$_2$ film. This analysis could provide us values of two important physical parameters, *i.e.* ($U + \Delta$) ~ 3.4 eV and $J_H$ ~ 0.7 eV. As the MIT occurs, we found that the spectral weight up to 5.0 eV becomes redistributed. Based on these findings, we will propose the possible electronic structure changes, especially near the Fermi level, and discuss their implications for the underlying MIT mechanisms.

## II. FABRICATIONS AND CHARACTERIZATIONS OF VO$_2$ FILM

High quality VO$_2$ thin films, with thickness of about 70 nm, were prepared on *R*-cut Al$_2$O$_3$ substrates using the pulsed laser deposition method. In the narrow range of deposition conditions (*i.e.* 55 - 60 mTorr of 0.9 Ar + 0.1 O$_2$ gas and 400 ℃ of substrate temperature), the fil-ms could be grown epitaxially with their c-axis normal to the substrate.[2,26] Values of dc-resistivity, $\rho(T)$, and c-axis lattice constants, $c_r(T)$, were obtained between 300 and 380 K using a conventional four-probe method and x-ray diffraction measurements, respectively.

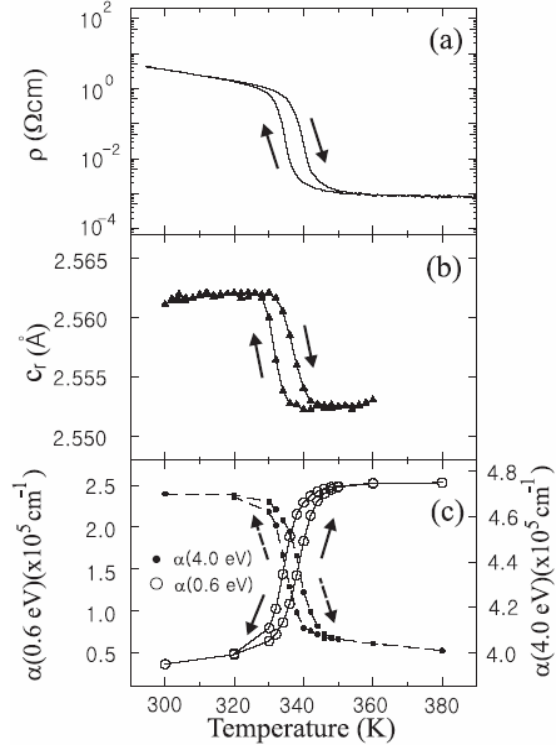

FIG. 1. (a) Temperature dependences of *dc*-resistivity , (b) $c_r$-axis lattice constant, and (c) absorption coefficient at 0.6 eV, $\alpha(0.6\ eV)$, and at 4.0 eV, $\alpha(4.0\ eV)$, of the $VO_2$ film. In (c), closed and open circles represent $\alpha(0.6\ eV)$ and $\alpha(4.0\ eV)$, respectively. Arrows guide the process of temperature changes.

Transmission spectra of a high quality $VO_2$ film were obtained in the photon energy range of 0.5 - 5.0 eV and in the temperature range of 100 - 380 K using a grating type spectrophotometer. Then, its $\alpha(\omega)$ were obtained by taking a logarithm of the transmittance and dividing it by the sample's thickness. Due to the first-order phase transition nature of the $VO_2$, $\alpha(\omega)$ were obtained during the warming and cooling runs. Near the phase transition temperature, especially near 340 K, the temperature was finely controlled with 2.0 K intervals with an accuracy of ±0.3 K. We found that the $\alpha(\omega)$ spectra at 300 K were nearly the same between the cooling and heating runs.

Figures 1(a) and 1(b) show the *T*-dependent $\rho(T)$ and $c_r(T)$ curves of the $VO_2$ film, respe-ctively. The arrows indicate the warming and the cooling runs. As $T$ increases, r($T$) abruptly decreases by three orders of magnitude near 340 K. The r($T$) curve shows a strong hysteresis. Accompanying the resistivity changes, the c$_r$($T$) curve also shows drastic changes at nearly the same temperature regions. The strong hysteresis curves imply that the structural phase transition from the monoclinic insulating to the tetragonal metallic phases should have a first-order transition nature, consistent with previous studies on VO$_2$ single crystals.[6]

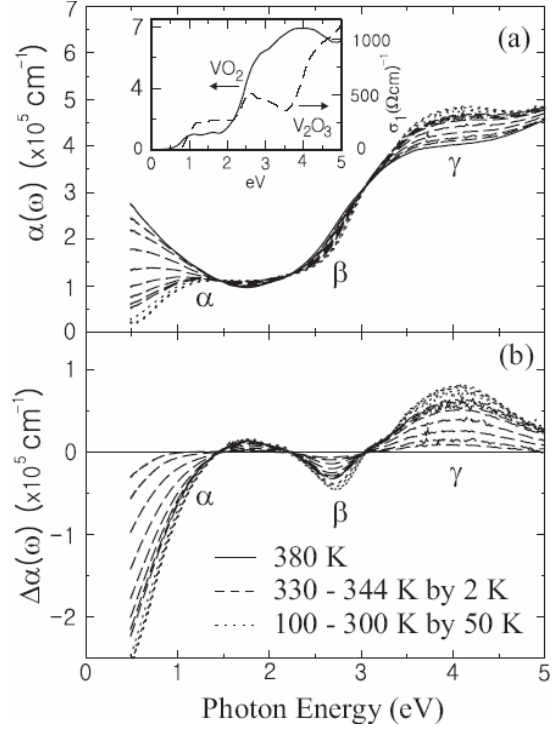

FIG. 2. (a) Temperature-dependent absorption coefficient spectra of the VO$_2$ film during the heating run, and (b) the amount of absorption coefficient change between the given temperature and 380 K ($\Delta\alpha(\omega) = \alpha(\omega)_T - \alpha(\omega)_{T=380\ K}$). The solid and dotted lines are the optical data in the metallic and insulating phases, respectively. The dashed lines are the optical data in the phase transition regions. The inset shows the absorption coefficient spectrum of VO$_2$ film (the solid line)[16] and the conductivity spectrum of single crystal V$_2$O$_3$ in the insulating phase (the dashed line),[36] respectively.

### III. TEMPERATURE DEPENDENT ABSORPTION SPECTRA OF THE VO$_2$ FILM

Figure 2(a) shows the $T$-dependent $\alpha(\omega)$ of the VO$_2$ film during the heating run. In the metallic phase, *e.g.* at 380 K, one can clearly observe a Drude-like peak below 1.0 eV and a strong peak around 4.0 eV. As *T* decreases down to the insulating phase, *e.g.* at 100 K, the Drude-like peak becomes strongly suppressed and forms an incoherent peak near 1.3 eV, and the peak around 4.0 eV is shifted to higher energy. Hereafter, we call the broad peak near 1.3 eV and the strong peak near 4.0 eV Peaks a and g, respectively. The optical gap was estimated to be around 0.6 eV, similar to the value reported in a previous study on single crystalline $VO_2$.[14] By integrating $\alpha(\omega)$ up to 5.0 eV, we found the total spectral weight under 5.0 eV remains almost independent of *T*, indicating that the electronic changes related to the MIT should occur near the Fermi energy.

In order to show the *T*-dependent changes in $\alpha(\omega)$ more clearly, we obtained the difference spectra by subtracting from each spectrum that at 380 K, *i.e.* $\Delta\alpha(\omega) = \alpha(\omega)_T - \alpha(\omega)_{T=380 K}$, and plotted them in Fig. 2(b). While the spectral weight of the Drude-like peak decreases with *T*, that of the Peak g increases. In $\Delta\alpha(\omega)$, one can clearly notice the appearance of another peak near 2.7 eV, which is not easily observable in $\alpha(\omega)$. [Compare Figs. 2(a) and 2(b).] The 2.7 eV peak shows little energy shift and increases its spectral weight with increasing *T*. In fact, the existence of the 2.7 eV peak has already been noticed by other workers using a thin film or a single crystal. [see the inset of Fig. 2(a).][15,16] Hereafter, we call the 2.7 eV peak Peak b.

## IV. PEAK ASSIGNMENTS BASED ON THE CONVENTIONAL BAND PICTURE
### A. The Band Diagram

Figures 3(a) and 3(b) show the most widely accepted band diagram[27] of $VO_2$ in the metallic and insulating phases, respectively. The designation of the bands involved was made following Goodenough's early notation.[4] In the metallic phase with the tetragonal rutile structure, the triply degenerate V $t_{2g}$ band is split into the $d_\parallel$ and the doubly degenerate $\pi^*$ bands due to the crystal field splitting $\Delta$ under the tetragonal structure. Note that both the V $d_\parallel$ and the V $\pi^*$ bands cross the Fermi level, and that an electron partially fills the $d_\parallel$ and $\pi^*$ bands. On the other hand, in the insulating phase with the monoclinic structure, the V $\pi^*$ band shifts upward by about 0.5 eV, resulting in an unoccupied band, and the V $d_\parallel$ band becomes split into one occupied and the other unoccupied bands. Note that the Fermi level should be located between the occupied V $d_\parallel$ and V $\pi^*$ bands. On the other hand, the filled O $2p$ band shows little change between the metallic and the insulating phases.[27,28]

### B. Analysis of $\alpha(\omega)$

Based on this schematic band diagram, the overall features of $\alpha(\omega)$ can be understood.

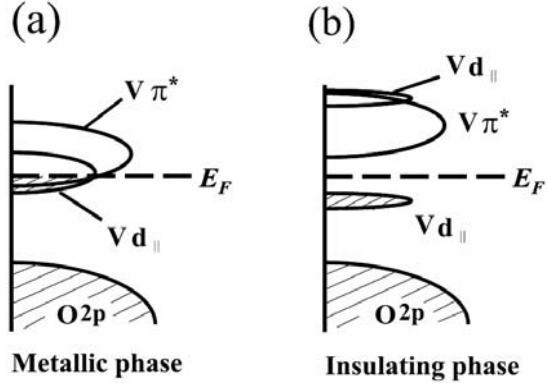

FIG. 3. Schematic band diagrams of VO$_2$ for (a) the metallic and (b) the insulating phases.[27]

In the low energy region, $\alpha(\omega)$ can be characterized by the strong Drude-like peak in the metallic phase and the optical gap in the insulating phase. The Drude-like peak should be attributed to the free carrier responses in the broad $d_\parallel$ and $\pi^*$ bands, and the optical gap should correspond to the energy difference between the upper edge of the occupied $d_\parallel$ band and the lower edge of the unoccupied $\pi^*$ band.

In the high energy region, $\alpha(\omega)$ can be characterized by the strong Peak $\gamma$, which experiences an energy shift of 0.5 eV to higher energy when the sample enters the insulating phase. Peak $\gamma$ can be attributed to the transition from the O 2p to the V $\pi^*$ bands. Note that the density of states for the $\pi^*$ band is about two times larger than that for the $d_\parallel$ band, since the $\pi^*$ band is doubly degenerate. Also, the $\pi^*$ band has a stronger hybridization with the O 2p band than the $d_\parallel$ band.[29] Therefore, it is quite reasonable to assign Peak $\gamma$ as the transition from the O 2p to the V $\pi^*$ bands. In addition, Peak $\gamma$ is found to shift upward by around 0.5 eV as $T$ decreases, which is consistent with the upward shift of the $\pi^*$ band. Since the unoccupied $d_\parallel$ band is located at a higher energy level than the $\pi^*$ band, the optical transition from the O 2p to the V $d_\parallel$ bands in the insulating phase should be located at a higher energy region than the Peak $\gamma$,[28] which cannot be observed in our measured energy region, *i.e.* 0.5 - 5.0 eV. [However, the earlier reflectance studies[28] claimed that the optical transition from the O 2p to the V $d_\parallel$ bands should have a peak around 5.0 eV.]

### C. Difficulties of Assigning the *d-d* Transitions in the Band Picture

Let us address the low energy region peaks, *i.e.* the Peaks $\alpha$ and $\beta$, in the insulating phase. From the band diagram, one can consider two *d-d* transitions: namely, from the occupied $d_\parallel$ to the unoccupied $\pi^*$ bands, and from the occupied $d_\parallel$ to the unoccupied $d_\parallel$ bands. Although the $d_\parallel$ and the $\pi^*$ orbitals are orthogonal to each other at the same V atom, each $d$ orbital in one site could have some hybridization with that at the neighboring site via the O $2p$ orbitals. Figures 4(a) and 4(b) show the crystal structure of VO$_2$ in the metallic phase and the configurational arrangements of the V $d_\parallel$, V $\pi^*$, and O $2p_x$ orbitals. [The x-y plane in Fig. 4(b) is marked as the box with the thick solid lines in Fig. 4(a)]. Note that the $\pi^*$ ($d_{x'z'}$) orbital at the corner site has a π-bonding with the O $2p_x$ orbital, and that the $d_\parallel$ ($d_{x^2-y^2}$) orbital at the center site also has a bonding with the same O $2p_x$ orbital. Due to this overlap through the O $2p_x$ orbital, the transition from the $d_\parallel$ to the $\pi^*$ bands could be allowed. In this respect, Peak α could be assigned as the d-d transition from the occupied $d_\parallel$ to the unoccupied $\pi^*$ bands. Also, Peak β should be assigned as the other d-d transition from the occupied $d_\parallel$ to the unoccupied $d_\parallel$ bands. These assignments are consistent with recent band calculations.[30]

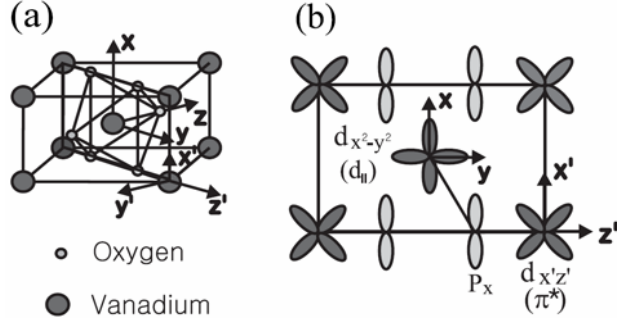

FIG. 4. (a) Tetragonal structure of VO$_2$ in the metallic phase. (b) The configurational arrangement of vanadium $d_\parallel$ and $\pi^*$ orbitals, and oxygen $p_x$ orbital. Thick line in (b) indicates the $d_\parallel$ - $\pi^*$ transition. The x-y plane in this diagram is marked as the box with thick solid lines in (a).

However, it is not correct to assign Peak β as the transition from the occupied $d_\parallel$ to the unoccupied $d_\parallel$ bands. As the phase transition occurs with decreasing T, Shin et al. showed using photoemission spectroscopy and reflectance spectra in the ultra-violet region that the unoccupied $d_\parallel$ band experiences a significant up-shift by about 1.1 eV.[28] If Peak β is assigned as the transition from the occupied $d_\parallel$ band to the unoccupied $d_\parallel$ one, it should move upward by about 1.1 eV, which is inconsistent with our observations. On the other hand, Verlur et al. assigned Peak β as the optical transition from the O $2p$ to the V $t_{2g}$ bands[15], but the strength of Peak β is too weak to be assigned as the charge transfer transition. Therefore, there seems to be some difficulties in properly assigning low energy region peaks in the insulating phase.

# V. MULTIPLE d-d PEAK ASSIGNMENT BASED ON THE ORBITALLY DEGENERATE HUBBARD MODEL

## A. The Orbitally Degenerate Hubbard Model (ODHM)

To check whether the two *d-d* transitions (*i.e.* Peaks α and β) should be understood in terms of the correlation induced peaks, we decided to apply the ODHM, which is an extension of the single-band Hubbard model by taking into account the orbital degeneracy.[20,21,22,25,31,32] To apply this model, let us first consider possible spin correlations of the nearest neighboring V $t_{2g}$ levels. In the insulating $VO_2$ phase, spins of the V atoms in one dimer should have an antiferromagnetic (AFM) spin arrangement, but the dimerization results in a nonmagnetic state.[33] In other words, the spins of the V atoms between the next neighboring dimers are known to have no spin correlation.[10,11] Therefore, we should think of two kinds of the nearest neighboring spin configurations: ferromagnetic (FM) and AFM states. Now, let us consider possible orbital correlations between the nearest neighboring atoms. Note that the $d_{\parallel}$ and the $\pi^*$ orbitals are split by $\Delta$, so one electron per V atom should occupy the $d_{\parallel}$ orbital.[34] When two electrons at the nearest neighbor sites occupy the same orbitals, as in our case, we will call it a ferro-orbital (FO) correlation. Consequently, we should consider two kinds of spin/orbital configurations for $VO_2$: namely, the FM/FO and the AFM/FO configurations.

Figure 5 shows schematic diagrams of possible optical transitions in the given configurations. In the FM/FO configuration, an optical transition from the $d_{\parallel}$ to the $\pi^*$ bands could be possible, but an optical transition from the $d_{\parallel}$ to the neighboring $d_{\parallel}$ bands should not be allowed due to the Pauli exclusion principle. For the former, the energy cost can be estimated by $U - 3J_H + \Delta$.[22] On the other hand, in the AFM/FO configuration, both optical transitions from the $d_{\parallel}$ to the $\pi^*$ bands and from the $d_{\parallel}$ to the $d_{\parallel}$ bands could be allowed. For the former and the latter, the energy cost can be estimated by $U - J_H + \Delta$ and $U + 2J_H$, respectively.[22] Note that Fig. 5 corresponds to an extension of the ODHM to the case where the energy levels of $t_{2g}$ electrons are split by the crystal field splitting $\Delta$.

## B. Analysis of α(ω)

Based on the ODHM, the Peaks α and β can be assigned to the two lowest-lying *d-d* transitions with excitation energies of $U - 3J_H + \Delta$ and $U - J_H + \Delta$, respectively. Since these peaks appear at 1.3 eV and 2.7 eV, the values of $U + \Delta$ and $J_H$ can be estimated to be 3.4 eV and 0.7 eV, respectively. Recently, Biermann *et al.* calculated the spectral function of $VO_2$ using the local density approximation and the cluster expansion of the dynamical mean field theory.[35] They found that a large value of $U$ (e.g., ≥ 5.0 eV) will lead to an insulating phase, but should display a large magnetic susceptibility, which is not consistent with experiments.[33] On the other hand, a small value of $U$ (e.g., ~ 2.0 eV) will stabilize an insulating phase of $VO_2$, only when $J_H$ is taken to be quite small. To understand the magnetic property and the stable insulating phase of $VO_2$, they argued that $U \sim 4.0$ eV and $J_H \sim 0.68$ eV. These theoretical values agree with our experimentally determined values reasonably well.

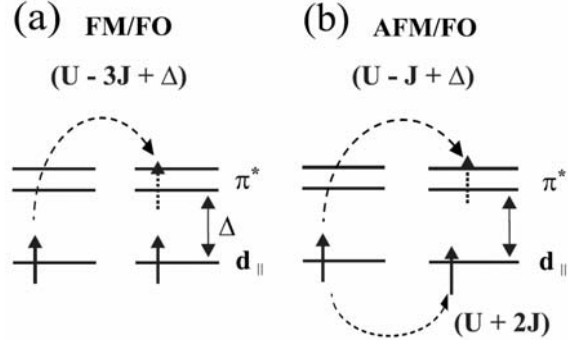

FIG. 5. The schematic diagrams of the $t_{2g}$ orbitals at two neighboring sites with (a) FM/FO and (b) AFM/FO configurations in the insulating $VO_2$ phase. FM, AFM, and FO represent ferromagnetic, anti-ferromagnetic, and ferro-orbital configurations, respectively. Possible optical transitions are shown as dotted arrows with the required energy costs in parentheses. Note that the schematic diagrams show the extension of the ODHM to the case where the $t_{2g}$ electron levels are split by the crystal field splitting $\Delta$.

In the ODHM, both of the correlation peaks below the charge transfer excitation come from the optical transitions from the $d_\parallel$ to the $\pi^*$ bands. Therefore, neither Peaks α nor β have to move with the MIT, so the difficulties of assigning the d-d transition in the band picture, which were described in Section IV.C, do not exist in the ODHM. Although both correlation peaks come from the transitions from the $d_\parallel$ to the $\pi^*$ bands, the different peak energy position is possible due to the orbital multiplicity effects, as demonstrated in Fig. 5. Therefore, the ODHM can explain the optical spectra of $VO_2$ better than the conventional band picture.

The dashed line in the inset of Fig. 2(a) shows the optical spectra of the $V_2O_3$ compound, reported by Thomas et al.[36] To further confirm our assignments for $VO_2$, we also applied the ODHM to explain the optical spectra of the other binary vanadium oxide. There are two correlation peaks around 1.2 eV and 3.0 eV in the conductivity spectra of $V_2O_3$.[36,37] For this compound, the ODHM predicts that the d-d transition should appear at $U - 3J_H$, $U$, and $U + 2J_H$.[22] The experimentally observed peaks should correspond to the two lower-lying two correlation peaks, *i.e.* $U - 3J_H$ and the $U$ peaks. Comparing these assignments with experimentally observed peak positions, we could obtain $U \sim 3.0$ eV and $J_H \sim 0.6$ eV for $V_2O_3$. Note that these values are also consistent with those of $VO_2$.

## VI. SPECTRAL WEIGHT CHANGES RELATED TO THE METAL-INSULATOR TRANSITION

### A. Spectral weight redistribution due to the electronic structure change

Note that Peak β also shows a discernable *T*-dependence and a hysteresis near MIT. Although such *T*-dependence could be attributed to the structural change, it is a little difficult to understand spectral weight change quantitatively. To elucidate its *T*-dependence by evaluating the matrix element of the electric dipole transition, we should take into account numerous factors, such as bond-distance, spin configuration, and orbital arrangement. However, to the best of our knowledge, it is not clear whether the bond-distance between V atoms in different dimer chains will become shorter or longer than before the MIT.[6] Because of the tilting of the dimer, the fine details of the orbital arrangement are not known yet. For this reason, more structural investigation seems to be highly desirable before we can understand the *T*-dependent spectral weight of Peak β quantitatively.

Figure 1(c) shows the *T*-dependences of $\alpha(\omega)$ at 0.6 eV (the solid circles) and at 4.0 eV (the open circles). The former and the latter should be closely related with the strengths of free carrier response and the *p-d* transition, respectively. Near 340 K, where MIT occurs, $\alpha(0.6$ eV$)$ and $\alpha(4.0$ eV$)$ experience abrupt changes and show strong hysteresis. It is interesting to note that such abrupt changes of $\alpha(0.6$ eV$)$ and $\alpha(4.0$ eV$)$ are quite similar to that associated with $c_r$, as shown in Fig. 1(b). This result clearly demonstrates that the electronic structural changes with the MIT should be closely related with the structural changes. On the other hand, as mentioned in Section III, the total spectral weight under 5.0 eV remains nearly independent of *T*. Due to the small value of Peak β, this sum rule analysis indicates that most of the spectral weight changes occur between the free carrier response and Peak γ. Especially, the abrupt spectral weight change of Peak γ should be closely related with the rearrangement of the $\pi^*$ band, *i.e.* the change of the hybridization strength between the O $2p_x$ and the V $\pi^*$ bands.[29] Therefore, the spectral weight changes might provide some insights on the underlying mechanisms of the MIT.

### B. Spectral weight redistribution due to the electron-lattice interaction and other factors

We discussed the role of the electron-electron correlation in our optical spectra of $VO_2$.

However, to describe the spectral features more precisely, other important physical parameters such as the electron-lattice interaction cannot be ignored. Recently, the electron-lattice interaction has been emphasized to explain the $T$-dependent spectral weight changes of the lowest peak in the photoemission spectra. Okazaki *et al.* have found that the $T$-dependent spectral weight change of the peak in the insulating phase cannot be understood only by the simple thermal broadening.[38] They have shown that the $T$-dependent spectral weight changes well agree with the predicted spectra obtained from a theoretical model based on the Frank-Condon mechanism. This kind of transition has been frequently observed in numerous molecules,[39] but it has been considered that such a transition could be strongly suppressed in a real solid system, since such an excitation can be easily delocalized. However, in the model, the authors argued that the Gaussian broadening of the pure electronic spectral function due to the Frank-Condon mechanism could occur in $VO_2$ due to the extremely strong electron–lattice coupling (*i.e.*, electron-phonon coupling constant g = 40) in this material.[38]

The absorption coefficient spectra, in Fig. 2(a), show a broadening near the optical gap and concomitantly a suppression of the intensity around the center of the peak α as $T$ increases in the insulating phase. The $T$-dependent spectra change might be able to be explained by the Frank-Condon mechanism.[40] However, it should be noted that such spectral change could not be a sufficient condition to invoke the electron-lattice interaction as a sole driving mechanism: in other words, it can be also explained by other mechanisms. One possibility is that it can be attributed to the defect and/or non-stoichiometric origins. We measured the absorption coefficient spectra of numerous $VO_2$ thin films which were grown under different oxygen partial pressure conditions. We observed that the optical gap became reduced to lower values when the films were grown under a low oxygen partial pressure. It was also observed that the $T$-dependence of $\alpha(\omega)$ near the optical gap for the films grown under low oxygen partial pressure was larger than that for the films grown under optimal oxygen partial pressure. These results indicate that the spectral weight change near the optical gap could be due to the extrinsic origins. It is well known that the defects can induce the similar $T$-dependence of optical spectra, the so-called 'Urbach tail' feature.[41] Another possible mechanism can be the coexistence of the metallic and the insulating phases near the phase transition boundary. As we showed recently,[26,42] the effective medium approximation could explain the $T$-dependent spectral weight change qualitatively in terms of the percolative transition. The percolative mixture of the metallic and the insulating phase was also recently confirmed by a scanning tunneling spectroscopy experiment.[43] Therefore, the $T$-dependent spectral weight change near the optical gap in the insulating phase could be attributed to an increase in the metallic region in the sample as $T$ approaches MIT. To understand the $T$- dependence of the spectra in the insulating phase precisely, more careful studies are highly desirable.

## VII. DISCUSSIONS ON THE DRIVING MECHANISM OF THE VO$_2$ METAL-INSULATOR TRANSITION

As we mentioned in the Introduction, two mechanisms have been proposed as the main driving forces of the MIT in VO$_2$, however, the elucidation of the basic MIT mechanism is still lacking. One mechanism is based on the Peierls' lattice instability during the phase transition and the other is based on strong electron correlation effects. Figs. 3(a) and (b) schematically show the two extreme ends of MIT, *i.e.* the metallic and the insulating phases, respectively. A simple way to compare these two mechanisms can be found by understanding what will be the driving force in changing the VO$_2$ electronic structures from Figs. 3(a) to (b).

According to the Peierls' lattice instability picture,[4,5,6,7,8] the spins in the vanadium ions will become dimerized during the tetragonal-monoclinic structural change. As shown in Fig. 4(b), the dimerization should occur along the c-axis. Under this structural change, the $d_{\parallel}$ band will be split into two bands and have a spin Peierls' gap. In this picture, the insulating gap in Fig. 3(b) should be mainly attributed to the dimerization. The shift of the $\pi^*$ band should be accompanied due to the shift of the Fermi surface by the gap opening in the $d_{\parallel}$ band. On the other hand, according to the strong electron correlation picture,[9,10,11,44] the $T$-dependent structural changes will make the $\pi^*$ band move upward. When the $\pi^*$ band becomes away from the Fermi energy, the $d_{\parallel}$ band will become half-filled. Due to the strong electron-electron correlation in the narrow $d_{\parallel}$ band, this band will be split into the lower and the upper Hubbard bands, resulting in a Mott gap in the insulating state. In this picture, the insulating gap in Fig. 3(b) should be mainly attributed to the Coulomb interactions between electrons.

Elucidating the origin of the insulating gap from the optical spectroscopy might not be simple. In addition, actual measurements on the $d_{\parallel}$-$d_{\parallel}$ transition peak, which corresponds to the $U + 2 J_H$ in the ODHM, seem to be difficult, since they should be located near the strong *p-d* transition, *i.e.* above 4.0 eV. In spite of these difficulties, our investigation shows that there should exist signatures of the correlation effects in the optical spectra of VO$_2$.

In Section VI, we showed that the large spectral weight redistribution in the energy region up to 5.0 eV could be understood in terms of the rearrangement of the $\pi^*$ orbital near the MIT. Although there are large changes in α(ω), the total sum of the spectral weight up to 5.0 eV does not change much at every *T*. This indicates that the increased coherent spectral weight below 1.0 eV mainly should come from the decreased spectral weight of the Peak γ, *i.e.* the transition from the O 2$p$ to the V $\pi^*$ bands. Namely, the renormalization of the $\pi^*$ band should play a crucial role in the MIT of the VO$_2$ compound. This renormalization should come from the change of hybridization strength between the O 2$p$ and the V $\pi^*$ bands, which results in the change of screening between electrons in the $d_\parallel$ bands. Note that, in the strong electron correlation picture, such renormalization of the $\pi^*$ band is assumed to occur before the opening of the band gap, consistent with our experimental observation.

The existence of Peak β in the metallic phase also suggests the importance of the strong electron-electron correlation effects in this compound. The band picture cannot explain the transition peak below the energy region of the charge transfer peak in the metallic phase. However, in our observed α(ω), Peak β does exist near 2.7 eV in the metallic phase. Such an optical transition peak could be explained by the correlation-induced peak, which has been observed in the optical conductivity of correlated metals, such as CaRuO$_3$, and CaVO$_3$.[45,46] Actually, in the Peierls' model, there is no appropriate way to explain such a correlation peak in the metallic phase. Recently, Okazaki *et al.* observed a correlation induced peak of VO$_2$ by using photoemission spectroscopy measurements.[38] The existence of the correlation-induced peak in the metallic phase of VO$_2$ indicates the importance of the electron correlation in the metallic VO$_2$.

Finally, as already mentioned in Section V.B, by applying the ODHM to our experimental α(ω), we could obtain the values of ($U + \Delta$) and $J_H$, which are 3.4 eV and 0.7 eV, respectively. These values are reasonably in agreement with those predicted by dynamical mean field theory.[35] This agreement also demonstrates that the orbital degeneracy should be an important parameter to be considered when the physics of VO$_2$ is discussed in terms of the electron-electron correlation.

**VIII. SUMMARY**

We investigated temperature-dependent absorption coefficient spectra α(ω) of VO$_2$ films in the wide energy range of 0.5 - 5.0 eV. To explain multiple peak structures of the *d-d* transitions in α(ω) of VO$_2$, we applied the orbitally degenerate Hubbard model, which explicitly takes into account the orbital multiplicity. Using this model, we could explain the correlation-induced peak in the metallic phase of VO$_2$. We also could obtain values of ($U + \Delta$) and $J_H$, which are consistent with values from other studies. Based on our detailed peak assignments, we could also show that the energy shift of the $\pi^*$ band by hybridization leads to the disappearance of the coherent Drude peak, which is a quite an important scenario in the picture which explains the metal-insulator transition in VO$_2$ based on the electron-electron correlation effects. Contrary to the earlier spectroscopic studies based on the band picture, our study clearly demonstrated that the optical spectra of $VO_2$ have the signature of the electron-electron correlation.

**Acknowledgements**

We would like to thank Jaejun Yu for useful discussions. This work was supported by the Ministry of Science and Technology through the Creative Research Initiative program, and by KOSEF through the Center for Strongly Correlated Materials Research.

[1] F. J. Morin, Phys. Rev. Lett. **3**, 34 (1959).
[2] H.-T. Kim, B. G. Chae, D. H. Youn, S. L. Maeng, G. Kim, K. Y. Kong, and Y. S. Lim, New J. Phys. **6**, 52 (2004).
[3] L. A. L. de Almeida, G. S. Deep, A. M. N. Lima, and H. Neff, Opt. Eng. **41**, 2582 (2002).
[4] J. B. Goodenough, Phys. Rev. **117**, 1442 (1960).
[5] R. Srivastava and L. L. Chase, Phys. Rev. Lett. **27**, 727 (1971).
[6] D. B. McWhan, M. Marezio, J. P. Remeika, and P. D. Dernier, Phys. Rev. B **10**, 490 (1974).
[7] M. Gupta, A. J. Freeman, and D. E. Ellis, Phys. Rev. B **16**, 3338 (1977).
[8] F. Gervais and W. Kress, Phys. Rev. B **31**, 4809 (1985).
[9] A. Zylbersztejn and N. F. Mott, Phys. Rev. B **11**, 4383 (1975).
[10] J. P. Pouget, H. Launois, J. P. D'Haenens, P. Merenda, and T. M. Rice, Phys. Rev. Lett. **35**, 873 (1975).
[11] J. P. Pouget, H. Launois, T. M. Rice, P. Dernier, A. Gossard, G. Villeneuve, and P. Hagenmuller, Phys. Rev. B **10**, 1801 (1974).
[12] A. Cavalleri, Cs. Tóth, C. W. Siders, and J. A. Squier, F. Ráksi, P. Forget, and J. C. Kieffer, Phys. Rev. Lett. **87**, 237401 (2001).
[13] A. Cavalleri, Th. Dekorsy, H. H. W. Chong, J. C. Kieffer, and R. W. Schoenlein, Phys. Rev. B **70**, 161102 (2004).
[14] L. Ladd and W. Paul, Solid State Commun. **7**, 425 (1969).
[15] H. W. Verleur, A. S. Barker, Jr., and C. N. Berglund, Phys. Rev. **172**, 788 (1968).
[16] A. Gavini and C. C. Y. Kwan, Phys. Rev. B **5**, 3138 (1972).
[17] S. L. Cooper, Structure and Bonding **98**, 161 (2001).
[18] T. Arima and Y. Tokura, J. Phys. Soc. Jpn. **66**, 778 (1995).
[19] Y. Taguchi, K. Ohgushi, and Y. Tokura, Phys. Rev. B **65**, 115102 (2002).
[20] M. W. Kim, Y. S. Lee, T. W. Noh, J. Yu, and Y. Moritomo, Phys. Rev. Lett. **92**, 027202 (2004).
[21] J. S. Lee, Y. S. Lee, T. W. Noh, S. -J. Oh, J. Yu, H. Fukazawa, and Y. Maeno, Phys. Rev. Lett. **89**, 257402 (2002).
[22] J. S. Lee, M. W. Kim, and T. W. Noh, New J. Phys. **7**, 147 (2005).
[23] M. W. Kim, P. Murugavel, S. Parashar, J. S. Lee, and T. W. Noh, New J. Phys. **6**, 156 (2004).
[24] N. N. Kovaleva, A. V. Boris, C. Bernhard, A. Kulakov, A. Pimenov, A. M. Balbashov, G. Khaliullin, and B. Keimer, Phys. Rev. Lett. **93**, 147204 (2004).
[25] M. W. Kim, S. J. Moon, J. H. Jung, Jaejun Yu, Sachin Parashar, P. Murugavel, and T. W. Noh, cond-mat/0504616 (2005).
[26] Y. J. Chang, C. H. Koo, J. S. Yang, Y. S. Kim, D. H. Kim, J. S. Lee, T. W. Noh, H. -T. Kim, and B. G. Chae, Thin Solid Films **486**, 46 (2005).
[27] H. Abe, M. Terauchi, M. Tanaka, S. Shin, and Y. Ueda, Jpn. J. Appl. Phys. **36**, 165 (1997).
[28] S. Shin, S. Suga, M. Taniguchi, M. Fujisawa, H. Kanzaki, A. Fujimori, H. Daimon, Y. Ueda, K. Kosuge, and S. Kachi, Phys. Rev. B **41**, 4993 (1990).
[29] V. Eyert, Ann. Phys. (Leipzig) **11**, 650 (2002).
[30] A. Continenza, S. Massidda, and M. Posternak, Phys. Rev. B **60**, 15699 (1999).
[31] S. Di Matteo, N. B. Perkins, and C. R. Natoli, Phys. Rev. B **65**, 054413 (2002).
[32] H. F. Pen, J. van den Brink, D. I. Khomskii, and G. A. Sawatzky, Phys. Rev. Lett. **78**, 1323 (1997).
[33] K. Kosuge, J. Phys. Soc. Jpn. **22**, 551 (1967).
[34] Here, we use Goodenough's early notation.[4] Under the tetragonal crystal field, the $t_{2g}$ orbitals will be split into $a_{1g}$ and $e_g^\pi$ states. The $d_\parallel$ and the $\pi^*$ orbitals correspond to the $a_{1g}$ and the $e_g^\pi$ orbitals, respectively.

[35] S. Biermann, A. Poteryaev, A. I. Lichtenstein, and A. Georges, Phys. Rev. Lett. **94**, 026404 (2005).
[36] G. A. Thomas, D. H. Rapkine, S. A. Carter, and A. J. Millis, Phy. Rev. Lett. **73**, 1529 (1994).
[37] M. J. Rozenberg, G. Kotliar, and H. Kajueter, Phys. Rev. B **54**, 8452 (1996).
[38] K. Okazaki, H. Wadati, A. Fujimori, M. Onoda, Y. Muraoka, and Z. Hiroi, Phys. Rev. B **69**, 165104 (2004).
[39] Mark Fox, *Optical Properties of Solid* (Oxford University Press, NewYork, 2001).
[40] Philip B. Allen and Vasili Perebeinos, Phys. Rev. Lett. **83**, 4828 (1999).
[41] Z. Yang, K. P. Homewood, M. S. Finey, M. A. Harry, and K. J. Reeson, J. Appl. Phys. **78**, 1958 (1995).
[42] H. S. Choi, J. S. Ahn, J. H. Jung, T. W. Noh and D. H. Kim, Phys. Rev. B **54**, 4621 (1996).
[43] Y. J. Chang, private communication.
[44] H.-T. Kim, B. G. Chae, D. H. Youn, G. Kim, K. Y. Kang, S. J. Lee, K. Kim, and Y. S. Lim, Appl. Phys. Lett. **86**, 242101 (2005).
[45] J. S. Lee, Y. S. Lee, T. W. Noh, K. Char, S.-J. Oh, J.-H. Park, C. B. Eom, T. Takeda, and R. Kanno, Phys. Rev. B **64**, 245107 (2001).
[46] H. F. Pen, M. Abbate, A. Fuijmori, Y. Tokura, H. Eisaki, S. Uchida, and G. A. Sawatzky, Phys. Rev. B **59**, 7422 (1999).